\documentclass[aps,twocolumn,showpacs,floats,superscriptaddress]{revtex4}
\usepackage{graphicx}
\usepackage{dcolumn}
\usepackage{bm}
\usepackage{color}
\usepackage{amssymb}

\begin{document}
%%%%%%%%%%%%%%%%%%%%%%%%%%%%%%%%%%%%%%
%%%%%%%%%%%%%%%%%%%%%%%%%
\author{M. Guglielmino}
\affiliation{Dipartimento di Fisica, Politecnico di Torino, Corso Duca degli Abruzzi 24, I-10129 Torino, Italy}
\author{V. Penna}
\affiliation{Dipartimento di Fisica and CNISM Unit\`a di Ricerca, Politecnico di Torino, Corso Duca degli Abruzzi 24, I-10129 Torino, Italy}
\author{B. Capogrosso-Sansone}
\affiliation{Institute for Theoretical Atomic, Molecular and Optical Physics,
Harvard-Smithsonian Center of Astrophysics, Cambridge, MA, 02138}

%%%%%%%%%%%%%%%%%%%%%%%%%
\title{Mott Insulator to Superfluid transition in Bose-Bose mixtures
in a two-dimensional lattice
}
%%%%%%%%%%%%%%%%%%%%%%%%%%%%%%%%%%%%%%

\begin{abstract}
We perform a numeric study (Worm algorithm Monte Carlo simulations) of ultracold
two-component bosons in two-dimensional  optical
lattices. We study how the Mott insulator to superfluid transition is affected
by the presence of 
a second superfluid bosonic species. We find that, at fixed interspecies
interaction, the upper and lower boundaries of the Mott
lobe are differently modified. The lower
boundary is strongly renormalized even for 
relatively low filling factor of the second component and
moderate (interspecies) interaction.
The upper boundary, instead, is affected only for large enough filling of the
second component. Whereas boundaries are renormalized we find evidence
of polaron-like excitations.
Our results are of interest for current experimental setups. 
\end{abstract}


\maketitle

%67.85.Hj	Bose-Einstein condensates in optical potentials
% 67.85.Fg	Multicomponent condensates; spinor condensates
% 67.85.-d	Ultracold gases, trapped gases

%%%%%%%%%%%%%%%%%%%%%%%%%%%%%%%%%%%%%%%%
%%%%%%%%%%%%%%%%%%%
%\textit{Introduction}   
In the last decade, a considerable amount of theoretical and experimental
research has been devoted to the objective of using ultracold lattice bosons and
fermions
to address many outstanding condensed matter problems via Hamiltonian modeling.
Single species (Bose) Hubbard models, first introduced for fermions to describe
electrons in solids and considered to be the minimal model for high
temperature superconductivity, can be experimentally realized with ultracold
atoms in optical lattices~\cite{Jaksch} and have been extensively studied~\cite{Bloch,Lewenstein}.
If a second component is introduced, new fascinating phenomena 
and exotic quantum phases which cannot be accessed with single
species atomic gases, become available.
Prominent examples include the possibility of realizing quantum magnetic phases~\cite{Kuklov_Svistunov,Demler_Lukin,Soyler}, 
engineering offsite interactions~\cite{Soyler,Bruderer,Fleischhauer1} which would likely lead to
supersolid states~\cite{Buchler, Le Hur}, studying disorder/impurities effects~\cite{Fleischhauer2,Pfannkuche}.
\\ \indent
One of the fundamental questions to address in two-component systems is how the
presence of a second species affects the Mott insulator (MI)-superfluid (SF)
transition.
While MI phases of single component bosonic systems have been well established
experimentally in one-~\cite{Esslinger1}, two-~\cite{Porto}, and three-dimensions~\cite{Greiner}, 
the multi-component case results more challenging to understand as it involves additional degrees
of freedom and hence a much richer physics. There have been a number of recent experiments on 
Bose-Bose~\cite{Minardi1,Minardi2, Schneble} and Fermi-Bose~\cite{Ospelkaus,Esslinger2,Bloch,
Bloch2} 
mixtures which have investigated how the visibility of one component is affected by the presence
of a second one upon varying relative densities of the two species and/or the interspecies
scattering length. The overall finding is that the presence of a second component reduces the
visibility of the first bosonic species. A number of theoretical explanations which include
multi-band model~\cite{Das Sarma}, self-trapping effect~\cite{Pfannkuche}, changes in the
chemical potential due to the presence of external harmonic confinement~\cite{Buonsante}, have been provided.
\\ \indent 
In the present work we consider a homogeneous system of two-component bosons in a
square lattice with repulsive interspecies interaction. We study how the MI-SF transition of
the majority component is affected by the presence of a minority superfluid component upon varying
the density of the latter and the interspecies interaction. This system can be realized by loading
optical lattices with two different atomic species \cite{Minardi1, Minardi2}, or the same atomic
species in two different internal energy states \cite{Schneble}.  Interspecies
interaction strength $U_{ab }$ can be tuned either via Feshbach resonance
or by changing the Wannier functions overlap (in the presence of state-dependent lattices). 
Intraspecies interactions $U_{a}$ and $U_{b}$ can also be tuned via Feshabch resonance.
If the temperature is low enough, the system is accurately described by the two-component
Bose-Hubbard Hamiltonian
\begin{eqnarray}
%{}
& & H=-t_a\! {\sum}_{<ij>}  a^+_i\, a_j \, 
-t_b  {\sum}_{<ij>}  b^+_i\, b_j \nonumber \\ 
& & +U_{ab} {\sum}_{i}  n^{(\!a)}_i  n^{(\!b)}_i \, 
   + \frac{1}{2}{\sum}_{i,\alpha} U_{\alpha}\; n^{(\!\alpha)}_i
( n^{(\!\alpha)}_i-1) \nonumber \\
& & -\mu_a\!{\sum}_{i}  n_i^{(\!a)} \, -\mu_b\!{\sum}_{i}  n_i^{(\!b)}
\, . 
\label{hamiltonian}
\end{eqnarray}
where $n^{(a)}_i= a^+_i a_i$ and $n^{(b)}_i= b^+_i b_i$, and
bosonic annihilation (creation) operators $a_i$ ($ a^+_i $), $b_i$ ($ b^+_i $) 
satisfy the standard commutators $[a_i, a^+_j] = \delta_{ij} = [b_i, b^+_j] $.
Parameters $t_a$, $t_b$ represent the hopping amplitudes, $\mu_a$, $\mu_b$ the chemical 
potentials for the two bosonic species $A$ and $B$, respectively. Index 
$\alpha\! =\! a, b$ identifies the two component. 
\\ \indent
The phase diagram of model~\ref{hamiltonian} is very reach. It includes several stable 
phases which vary from double MI's~\cite{Iskin}, two independent SF's, and a mixture of
one MI and one SF, to less trivial phases like supercounterflow and checkerboard
solid~\cite{Kuklov_Svistunov, Demler_Lukin}.
In the following we present the first accurate results, based on path integral
Monte Carlo simulations by the Worm algorithm~\cite{WormA}, for the MI-SF
phase diagram of the majority component A. 
We study how the transition is affected by the presence of the minority SF component B
upon varying the density of the latter and the interspecies interaction. 
\\ \indent
Our results can be summarized as follows. We find that the net effect of the SF minority 
component B is to screen the effective onsite repulsion of component A thus resulting in a partial
suppression of the Mott region, 
with the upper and lower boundaries of Mott lobes being very differently renormalized.
Although the extension of the SF region
would imply an increase of the system coherence and seems to contradict experimental findings, we point out that, in order 
to achieve a conclusive understanding, further investigation 
on, e. g., effects of the second component on the superfluid state of the majority species and of
the presence of the harmonic confinement is needed. 
Finally, we find that, whereas boundaries are renormalized, particle and
hole elementary excitations get dressed. The decay in imaginary time of the 
momentum--space Matsubara Green function of the component in the MI is
reminiscent of the decay of the polaron Green function studied in the context of
an electron coupled to optical phonons (see e. g.~\cite{polaron_DMC}). 
%%%%%%%%%%%%%%%%%%%%%%%%%%%%%%%%%%%%%%%%%%%%%%%%%%%%%%%%%%%%%%%%%%%%%%%%%%%%%%%
%\section{Results}
\\ \indent 
%\textit{Results}.  
The large number of independent parameters occurring in Hamiltonian (\ref{hamiltonian})
involves an extremely structured phase diagram. 
In addition to the five parameters $t_\alpha$, $U_\alpha$ and $U_{ab}$ one must include
the boson number of each species ${\cal N}_a$ and ${\cal N}_b$,
controlled by chemical potentials $\mu_a$ and $\mu_b$. 
In the following we focus our attention on the first Mott lobe of species A, i.e.
filling factor $n_a= {\cal N}_a/M =1$, where $M$ is the number of lattice sites. 
We use the MI regime of the single-component system, described by Hamiltonian
(\ref{hamiltonian}) for ${\cal N}_b =0$, as the
reference case, and start our analysis by adding to the system
a small amount of componen B (${\cal N}_b << {\cal N}_a$) interacting with species A  
via a relatively weak interaction term $U_{ab}$. 
We then proceed by progressively increase either ${\cal N}_b $ or $U_{ab}$, while keeping fixed 
$U_b=10\, t_b$, and $t_b=t_a$ (in the following we choose $t_a$ as energy unit).
In order to obtain boundaries of the lobes, we either utilize the zero momentum 
Green function (GF) of component A (see~\cite{Single3D} for details), 
or study $n_a(\mu_{a})$ curves at fixed $n_b$. The latter method is used whenever the imaginary time decay of the 
GF differs from what expected for quasiparticle/hole elementary excitation (see below). 
\\ 
\indent
%%%%%%%%%%%%%%%%%%%%%%%%%%%%%%%%%%%%%%%%
\begin{figure}%
\begin{center}
\includegraphics[width=\columnwidth]{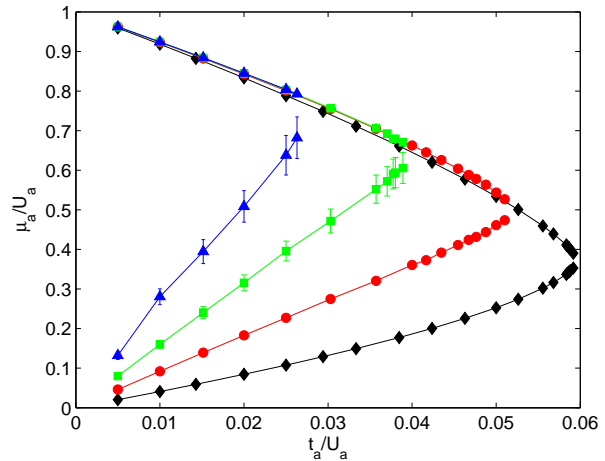}%
\end{center}
\caption{(Color online) First Mott lobe of species A with parameters $U_{b}=10\, t_a$, 
$t_b=t_a$, $n_b=0.1$ and varying $U_{ab} =10$, 15, 20$t_a$, circles, squares, triangles respectively. 
The lobe for the single-species case, i.e. ${\cal N}_b =0$,  is also plotted for reference (diamonds), with data taken from~\cite{Single2D}. 
Errorbars are within symbol size when not visible. Lines are to guide the eye.}
\label{fig:diffU12}%
\end{figure}
%%%%%%%%%%%%%%%%%%%%%%%%%%%%%%%%%%%%%%%%%
%
While in the absence of the second component the lobe boundaries are uniquely determined 
by varying parameters $\mu_a/U_{a}$ and $t_a/U_{a}$, in the presence of the second species,
the lobe boundaries are affected in different ways depending on the interplay of $U_{ab}$ with
parameters of species A.
We choose to use $t_a$ as our energy scale and determine the boundaries as $U_{a}$ changes, while
$U_{ab}$ is kept fixed for each lobe. We therefore expect the boundaries  of the new lobe to
coincide with those of the single-species case, i.e. $U_{ab} = 0$ or ${\cal N}_b =0$, in the
limit $t_a/U_{a} \to 0$, since this limit also implies $U_{ab}/U_{a} \to 0$.
We have performed simulations for linear system sizes $L =$ 10, 20, 30, temperature $T=t_a/2L$ and have found no
size effects within errorbars. The critical region at the tip of the lobes has not been studied
as it is beyond the scope of this work.
\\ \indent
In Fig.~\ref{fig:diffU12} we show results referring to fixed $n_b=0.1$ and 
$U_{ab}= \,10,\, 15,\, 20t_a$ corresponding to circles, squares and triangles, respectively. 
For comparison we also plot the single-species lobe (diamonds), taken from Ref~\cite{Single2D}. As expected, all lobes  
tend to overlap to the single component one in the $t_a/U_{a}\rightarrow 0$ limit.
In general, the change in shape of lobes involves a dramatic reduction of their extension even at
low density of component B considered here, $n_b=0.1$.
The suppression of the lobe becomes more pronounced as the interspecies interaction is increased. 
The presence of species B therefore inhibits the formation of the species--A Mott state by favouring delocalization, resulting into a critical point $t_a/U_{a}$ renormalized to lower values.
\\ \indent
Interestingly, while the lower boundary is strongly renormalized by the presence of component B,
the upper boundary is essentially indistinguishable from the single-species case.
Moreover, simulations done with $U_{ab}$ as large as $\sim 40 \, t_a$ suggest that, at $n_b\sim 0.1$, 
phase separation will occur before a sizeable shift in the upper boundary is detected. 
Indeed, in order to see any effect on the upper boundary, one has to go to larger density of 
component B (see Fig.~\ref{fig:diffn2} below).
A simple physical argument can be made by considering the case where component B is reduced to just one particle. 
The energy, counted from chemical potential $\mu_a$, required to create a hole excitation is reduced with respect to the
single-component case as the hole represents a chance for the system to lower its energy: 
the empty site can be occupied by particle B. As a consequence, the interaction 
with species A is fundamentally turned off.
The same effect cannot take place when a particle A is added to the lattice. In this case,
due to the large dilution of species B which allows the added particle A to avoid particle B, 
the energy cost for such an excitation remains substantially identical to that required when $n_b =0$.
%\\ \indent
These considerations can be extended to the case of $\mathcal N_b << \mathcal N_a$ and supported by the subsequent 
simple calculation. Denoting by $E({\cal N}_a, {\cal N}_b)$ the ground state energy of 
Hamiltonian (\ref{hamiltonian}) for negligible
$t_a$ and $t_b$, one finds the same particle-excitation energy gap 
$\Delta_{+} = E({\cal N}_a +1, {\cal N}_b) -E({\cal N}_a, {\cal N}_b) = U_a- \mu_a$ for both
${\cal N}_b =0$ and ${\cal N}_b \ne 0$. Conversely, 
in the case of hole-excitation, 
where $\Delta_{-} =E({\cal N}_a -1, {\cal N}_b) -E({\cal N}_a, {\cal N}_b)$, one finds 
$\Delta_{-} = \mu_a$ for ${\cal N}_b =0$, and $\Delta_{-} = \mu_a - U_{ab} < \mu_a$ for ${\cal N}_b \ne 0$. 
If we refer to Fig.~\ref{fig:diffU12} (and Fig.~\ref{fig:diffn2}), with model parameters rescaled
by $1/U_a$, we can observe that, for ${\cal N}_b\ne 0 $,
$\Delta_{-}/U_a \to ( \Delta_{-}/U_a)_{{\cal N}_b=0}$ 
when $U_a\to\infty$, confirming that different lobes will overlap in the limit $t_a/U_a\rightarrow 0$.
We can conclude that, at low B densities, the formation of a hole excitation is
manifestly favoured by the presence of species B, 
whereas the mechanism for creating a particle excitation is essentially unchanged.
%%%%%%%%%%%%%%%%%%%%%%%%%%%%%%%%%%%%%%
\begin{figure}
\begin{center}
\includegraphics[width=\columnwidth]{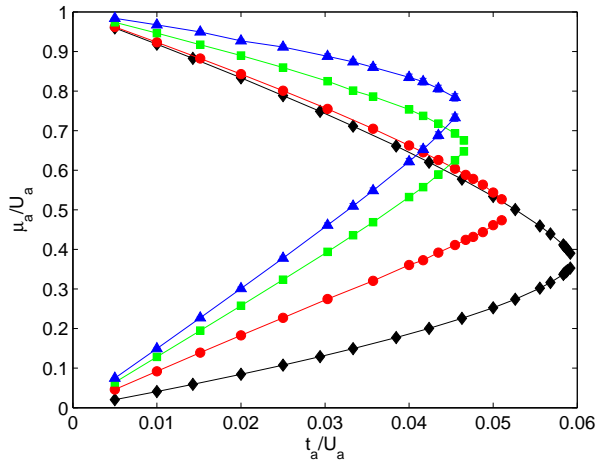}%
\end{center}
\caption{(Color online)
First Mott lobe of species A with parameters $U_{b}=10\, t_a$, 
$t_b=t_a$, $U_{ab}=10\, t_a$ and varying $n_b=0.1$, 0.5, 0.75, circles, squares, triangles respectively.  
The lobe for the single-species case, i.e. ${\cal N}_b =0$,  is also plotted for reference (diamonds), with data taken from~\cite{Single2D}
Errorbars are within symbol size when not visible. Lines are to guide the eye.}
\label{fig:diffn2}
\end{figure}
%%%%%%%%%%%%%%%%%%%%%%%%%%%%%%%%%%%%%%%%%%%
\\ \indent
Let us now turn to the discussion of the case of fixed $U_{ab} = 10 \, t_a$ 
and varying $n_b$. Fig.~\ref{fig:diffn2} shows the resulting Mott lobes for 
$n_b=0.1$, $0.5$ and  $0.75$, circles, squares and triangles, respectively. The single component case is also plotted for reference (diamonds).
As $n_b$ is increased the lobe size is further reduced, although, even at large $n_b=0.75$, the effect is 
less pronounced than what observed at $n_b=0.1$ and large $U_{ab}\sim 15-20\, t_a$.
The displacement of upper boundaries, though, becomes unequivocal for $n_b=0.5$ and 0.75.
Compared to the single component case, the energy 
required to create a particle excitation, counted from chemical potential $\mu_a$, significantly increases due to a 
large enough density of component B. The latter
discourages double occupancy of species A due to interaction $U_{ab}$ . 
%%%%%%%%%%%%%%%%%%%%%%%%%%%%%%%%%%%%%%%%%%%%%%%%%%%%%%%%
\begin{figure}
\begin{center}
\includegraphics[width=3in, angle=0]{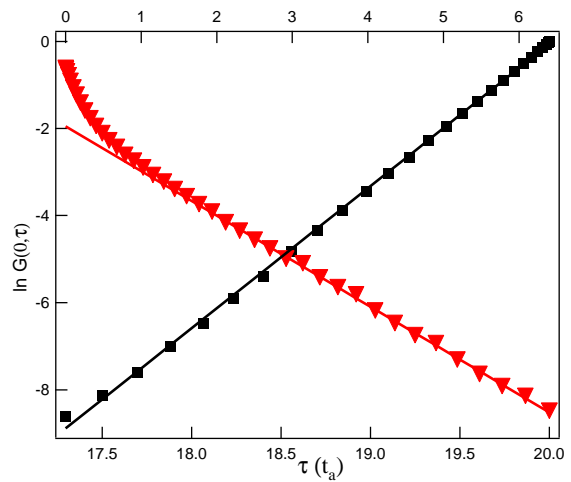}
\end{center}
\caption{
(Color online) Imaginary--time decay of the zero--momentum GF 
of component A for $L=10\, T=t_a/20$ $U_a=100\, t_a$, $U_b=10\, t_a$, $t_b=t_a$ \,
$n_b=0.1$, $U_{ab} = 10\, t_a$, $\mu_a = 10 \, t_a$ (triangles, lower $x$ axis)
and $\mu_a = 89\, t_a$ (square, upper $x$ axis) for hole and particle excitations,
respectively. Solid lines are a linear fit. The linear decay of $\ln G(0,\tau)$ on the hole excitation 
side is recovered only at later imaginary times, implying $Z_-<1$ (polaron-like behavior).
On the particle side, instead, the expected linear
behavior is observed.}
\label{fig:GF}
\end{figure}
%%%%%%%%%%%%%%%%%%%%%%%%%%%%%%%%%%%%%%%%%%%%%%%%
\\ \indent
In this scenario, a discussion of the imaginary time decay of the momentum space GF is appropriate.
The GF is defined as follows:
\begin{equation}
G(i,\tau)=\left\langle \mathcal T_\tau a_i^\dag(\tau) a_0(0)\right\rangle\; ,
\label{eq:GF}
\end{equation}
where $\tau$ is the imaginary time, and $\mathcal T_\tau$ represents the time ordering
operator. Using Lehman expansion and extrapolation 
to the $\tau\rightarrow\pm\infty$  limit one readily finds:
\begin{equation}
G(\mathbf p,\tau)\to Z_\pm \, e^{\mp\epsilon_\pm(\mathbf p)\tau},\qquad \tau\to\pm\infty.
\label{eq:explaw}
\end{equation}
The two limits describe single-particle/hole excitations in the MI
phase. Here $Z_{\pm}$ and $\epsilon_{\pm}$ are the
particle/hole spectral weight (or $Z$-factors) and energy,
respectively. In the grand canonical ensemble, excitation energies
are measured relative to the chemical potential. Chemical potentials $\mu_{\pm}$ for which the
energy gap for creating the particle/hole excitation with
$\mathbf{p}=0$ vanishes, are then retrieved by the exponential decay of $G( 0,\tau)$.
If elementary excitations are interpreted as polarons,
the spectral weights $Z_{\pm}$  carry information on
which fraction of the bare-(quasi)particle 
state is present in the polaron eigenstate. In the polaron GF, the exponential 
decay is expected to be recovered only at later imaginary times, therefore,
(properly normalized) spectral weights are smaller than 1, $Z_{\pm}<1$~\cite{polaron_DMC}.
This is precisely what we have observed. In Fig.~\ref{fig:GF} we plot $G_A(0,\tau)$
in log scale, for $U_a\!=\!100t_a$, $U_b\!=\!10t_a$, $t_b\!=\!t_a$, $n_b\!=\!0.1$, $U_{ab}\!=\!10\, t_a$, 
$\mu_a\! =\!10\, t_a $ and $\mu_a \!=\!89\, t_a$ 
corresponding to hole (triangles) and particle (squares) excitations respectively. 
By using the periodicity of GF in $\tau$, 
we show the plot for positive values $\tau \! \in \! [0, \beta]$, $\beta=1/T$.
Since we also perform the transformation $\tau \to -\tau$, the $\tau$ interval relevant 
to hole (particle) excitations is that starting from $\tau \! = \! 0$ (ending at $\tau \! = \! \beta$).
Solid lines are linear fits, with the \textit{y}-intersect 
providing spectral weights $Z_{\pm}$. 
Clearly, the decay on the hole excitation side initially differs from the purely exponential one expected. 
The latter is recovered at later imaginary times, implying $Z_-<1$. 
This is the typical behavior of the polaron GF. On the particles side, instead, the expected exponential decay is observed 
(here the lobe boundary coincides with the single component case).
We notice that, wherever lobe
boundaries differ from the single species case, we \textit{always} observe a polaron-like
behavior of the GF, both for particle and hole excitations. We interpret this observation as dressing of
elementary excitations, which we ascribe to the coupling of quasiparticles and holes of species A
with quasimomentum modes of SF species B. The latter play the role of phonon modes within the
tight-binding picture of Fr\"ohlich's model~\cite{FRO}. A simple argument seems to confirm this picture. 
Calling ${\bf R}_j$ the position of $j$-th site and expressing species--B operators in terms
of momentum modes 
$
b_j = {\sum }_{\bf k} b_{\bf k} \exp ({i {\bf k} \cdot {\bf R}_j} )/ \sqrt M
$ 
we can think the zero-momentum mode to be macroscopically occupied in the SF regime, 
which allows us to make use of the Bogoliubov approximation $b_0=b_0^\dag\simeq\sqrt{\mathcal N_b}$. 
Hamiltonian (\ref{hamiltonian}) reduces to
$H=H_U-\tilde\mu_a N_a-\mu_b N_b+ H_{pol}$, where the $H_U$ collects all the intraspecies interaction terms,
and $\tilde	\mu_a=\mu_a-U_{ab}\, n_b$ is a renormalized chemical potential for species A. $H_{pol}$ reads
\begin{eqnarray}
&&H_{pol}= 
-t_a\! {\sum}_{<ij>}  a^\dag_i\, a_j \, -  {\sum}_{\bf k} \omega_{\bf k}  {b^\dag}_{\bf k}\, b_{\bf k} \nonumber \\ 
&&+\tilde U_{ab} {\sum}_{j,{\bf k}} a^\dag_i\, a_j \, e^{i{\bf k}\cdot{\bf R}_j} \left(b_{\bf k} 
+ {b^\dag}_{-\bf k}\right), 
\label{eq:polaronic}
\end{eqnarray}
%$$
%+ H_U -\mu_a\!{\sum}_{i}  n_i^{(\!a)} \, -\mu_b\!{\sum}_{i}  n_i^{(\!b)}
%$$
%
%corretta il 22-6-10 where $\omega_{\bf k}= 2t_b \, \cos(d(k_x+k_y))$ 
where $\omega_{\bf k}= 2t_b \, [\cos(d k_x)+\cos(dk_y)]$ 
is the dispersion relation for species--B atoms, 
$d$ is the lattice spacing, and $\tilde U_{ab}=U_{ab}\sqrt{\mathcal N_b}/M$. 
The last term in~(\ref{eq:polaronic}) is of particular interest since it reproduces the interaction
scheme between phonons and electrons in the conductive band leading to
polaron excitations~\cite{FRO}. We notice that polaron physics in the context of impurities embedded in ultracold atomic 
systems has been a topic of great interests in the last few years, both experimentally and theoretically (see e.g.~\cite{Bruderer,polaron_DMC,polaron,Schneble}). In the present work, though, polaron physics 
emerge not from dressing of impurities but of elementary excitations.
%%%%%%%%%%%%%%%%%%%%%%%%%%%%%%%%%%%%%%%%%%%%%%%%%%%%%%%%%%%%%%%%%%%%%%%%%%%
\\ \indent 
Concluding, we have performed a numeric study of the MI-SF transition in the presence of a 
second SF bosonic species showing that, at fixed interspecies interaction, the upper and
lower boundaries of the Mott lobe are very differently modified. 
%While the lower boundary is strongly renormalized
%even for relatively low filling factor of the second component and moderate (interspecies)
%interaction, the upper boundary, is affected only for large enough filling of the second component.
The overall effect of the second component is reducing the extension of the MI lobe. 
We have also found that, whereas boundaries are renormalized, the decay of the GF
exhibits a polaron-like behavior. In the future it would be interesting to study 
how the second component affects the superfluidity of the majority component in the presence
of the harmonic confinement, as directly relevant to experiments. 
The investigation of the effective mass and the dispersion relation of polaronic excitations at strong,
weak and intermediate coupling regimes also deserves further study. 
These aspects will be considered elsewhere~\cite{WIP}.
%Another aspect deserving further investigation is the analysis on the effective mass 
%and the dispersion relation of polaronic excitations at strong,
%weak and intermediate coupling regimes~\cite{WIP}.
\\ \\
B. Capogrosso-Sansone (BCS) would like to thank D. Schneble, D. Pertot, B. Gadway, N. Prokof'ev, L. Pollet
for interesting discussion. V. Penna would like to thank F. Minardi for useful comments on the realization
of bosonic mixtures. The work of BCS was supported by the Insitute for Atomic, Molecular and Optical Physics (ITAMP).
%
%%%%%%%%%%%%%%%%%%%%%%%%%%%%%%%%%%%%%%%%%%%%%%%%%%%%%%%%%%%%%%%%

\end{document}